# Robust Self-Supervised Learning of Deterministic Errors in Single-Plane (Monoplanar) and Dual-Plane (Biplanar) X-ray Fluoroscopy

Jacky C.K. Chow, Steven K. Boyd, Derek D. Lichti, Janet L. Ronsky

*Abstract*—Fluoroscopic imaging that captures X-ray images at video framerates is advantageous for guiding catheter insertions by vascular surgeons and interventional radiologists. Visualizing the dynamical movements non-invasively allows complex surgical procedures to be performed with less trauma to the patient. To improve surgical precision, endovascular procedures can benefit from more accurate fluoroscopy data via calibration. This paper presents a robust self-calibration algorithm suitable for single-plane and dual-plane fluoroscopy.

A three-dimensional (3D) target field was imaged by the fluoroscope in a strong geometric network configuration. The unknown 3D positions of targets and the fluoroscope pose were estimated simultaneously by maximizing the likelihood of the Student-t probability distribution function. A smoothed k-nearest-neighbour (kNN) regression is then used to model the deterministic component of the image reprojection error of the robust bundle adjustment. The Maximum Likelihood Estimation step and the kNN regression step are then repeated iteratively until convergence.

Four different error modeling schemes were compared while varying the quantity of training images. It was found that using a smoothed kNN regression can automatically model the systematic errors in fluoroscopy with similar accuracy as a human expert using a small training dataset. When all training images were used, the 3D mapping error was reduced from 0.61—0.83 mm to 0.04 mm post-calibration (94.2—95.7% improvement), and the 2D reprojection error was reduced from 1.17—1.31 to 0.20—0.21 pixels (83.2—83.8% improvement). When using biplanar fluoroscopy, the 3D measurement accuracy of the system improved from 0.60 mm to 0.32 mm (47.2% improvement).

## I. INTRODUCTION

THORACIC and abdominal aortic aneurysm is a common condition associated with a high mortality rate. Half the patients with a ruptured aortic aneurysm die before arriving at the hospital, and less than 50% of the patients that reach the hospital survive [1], [2]. Elective repair of aortic aneurysm can reduce the morbidity and mortality in asymptomatic high-risk individuals [3]–[5]. Instead of an open surgery where the aorta is accessed through the chest wall for example, minimally-invasive surgery such as endovascular aneurysm repair (EVAR) allows surgeons to access the aorta by making a small incision in the groin and placing a stent graft in the aorta via a catheter. This has been shown to shorten both operating time and recovery time of the patient [6]. In order to guide the catheter insertion and the installation of stents in the operating room, X-ray fluoroscopy with radiopaque contrast agent is used intraoperatively by the vascular surgeon to visualize the structure of blood vessels and endovascular position of the equipment in real-time.

While the radiation dose is usually quite low (mean DAP 4.8 mGy.m2 over 21.8 min [7]), in complex endovascular procedures fluoroscopy may be on for hours leading to increased risk of radiation-induced cancer and tissue damage. Furthermore, the administered contrast is potentially nephrotoxic to patients with renal failure and/or diabetes. To address the above issues (i.e. reducing fluoroscopy time and the amount of contrast), image fusion of preoperative 3D computed tomography angiogram (CTA) or magnetic resonance angiogram (MRA) with intraoperative fluoroscopy can be utilized [8]–[10]. As noted in [11] this image fusion step can be performed as either 2D-3D registration or 3D-3D registration. While 3D-3D registration is more accurate, it requires a biplanar fluoroscopy. Regardless of the dimensionality of the fluoroscopic data, image fusion requires accurate image registration, which is better achieved if the individual image sources are free of distortions (i.e. are properly calibrated).

Distortion-free fluoroscopic images will become even more crucial with the growing adoption of robotic catheter steering systems. Endovascular robots can not only reduce radiation exposure to interventional radiologists, but further increase accuracy of stent placement and decrease contrast delivery [12]. The navigation accuracy of such a robotic system will greatly depend on the accuracy of the fluoroscopic imaging guidance system.

This article builds on a previous publication, [13], where a new automatic X-ray calibration method was proposed. Improvements to the geometric accuracy of fluoroscopic imaging systems was made by separating the residuals in a photogrammetric bundle adjustment into systematic and random components, and then modelling the systematic errors using a non-parametric machine learning method, the k-nearest-neighbour (kNN) regression. The modelling accuracy and behavior of kNN is studied in this paper by using four different error modelling schemes (described in Section VI). The novelty of the proposed method is that it is manufacturer independent and the X-ray fluoroscopy can be calibrated on-site, as regularly as needed, by any X-ray operator/technician with minimal additional training.

Manuscript submitted: August 9, 2019
J. C. K. Chow is with the Department of Medicine, Cumming School of Medicine, University of Calgary (email: jckchow@ucalgary.ca)
S. K. Boyd is with the Department of Radiology, Cumming School of Medicine, University of Calgary (email: skboyd@ucalgary.ca)
D. D. Lichti is with the Department of Geomatics Engineering, Schulich School of Engineering, University of Calgary (e-mail: ddlichti@ucalgary.ca).
J. L. Ronsky is with the Department of Mechanical and Manufacturing Engineering, Schulich School of Engineering, University of Calgary (email: jlronsky@ucalgary.ca)

The primary research question being addressed in this paper is whether the proposed automatic self-tuning calibration method can produce geometric accuracies comparable to the manual solution described in [14], where the calibration model was fine-tuned by an experienced photogrammetrist. This is beneficial because, by removing the need for an expert's knowledge, it can reduce the production and maintenance cost of such systems, thereby lessening the financial burden to public health care systems like the ones in Canada. The secondary purpose is to determine how much data is needed to model the systematic errors sufficiently using kNN. For data-driven modelling methods, it is often expected that the efficacy is improved with increasing amounts of data. However, there are financial costs involved with data capture: namely, the time of the X-ray technician and the time that the machine cannot be used in clinic. There are incentives for achieving an accurate calibration with reduced data (i.e. by using less time to acquire and process the data); for example, more imaging time can be allocated to patients, or the fluoroscope can be calibrated more frequently to ensure its optimal performance ([15] suggested that fluoroscopic systems may exhibit submillimeter 3D changes in measurement error over the course of a day).

This article provides a review of fluoroscopy calibration techniques (Section II) to highlight some limitations that will be addressed by the proposed method (Section III). This is followed by the details of the proposed mathematical algorithm in Section IV, a brief description of the proposed calibration procedure in Section V, results and analysis from the calibration of a mono- and dual-fluoroscopic imaging systems in Section VI, and conclusion in Section VII.

## II. Background

Fluoroscopic imaging systems such as the one considered in this article consist of many components: X-ray source, image intensifier, optics, video camera, and more. Any one of these components is prone to have manufacturing flaws resulting in distortions that can deteriorate the interpretability and accuracy of the final images contributing to error sources. For example, manufacturing imperfection of the image intensifier can cause artifacts such as lag, vignetting, veiling glare, pincushion distortion (i.e. a sub-type of radial lens distortion), sigmoidal distortion, and local distortions [16], [17]. While not all these artifacts can be easily calibrated (e.g. vignetting and veiling glare), others are more deterministic in nature (e.g. radial lens distortion, sigmoidal distortion, and local distortions) and are often considered in the error modelling process [18]–[20].

In [21], the calibration of a bi-planar fluoroscope was described as a three-step process: 1) correction of image distortion (i.e. solving for the additional parameters; AP); 2) calibration of the focus position (i.e. solving for the interior orientation parameters; IOP); and 3) solving for the relative orientation and position between the two fluoroscopic systems (i.e. the relative orientation parameters; ROP, which is closely related to solving the exterior orientation parameters, EOP, of an individual fluoroscopic system). In practice, steps 2) and 3) are often combined, resulting in a two-step procedure: 1) correction of image distortion; and 2) 3D space calibration [22]–[24]. In recent years, all three steps are combined into a unified optimization process to account for the parameter correlations that exist between the steps [14], [15], [25]. While there are certain benefits of doing a single optimization (e.g. accounting for correlations between intermediate parameters), when considering the calibration of fluoroscopic imaging systems, it is simpler to assess it in terms of its components, i.e. solving for AP, IOP, and EOP (and/or ROP).

The first step, removal of image distortion, is often divided into two categories: global methods [14], [18]–[20], [26], [27], where a polynomial is fitted to all the points in the image, or local methods [13], [17], [22], [28], where a more regional averaging or polynomial fitting is performed (e.g. local weighted mean, local unwarping polynomial). The choice of global or local methods has been an ongoing debate. It is considered that local methods can model local distortions better, but if the magnitude of this is small, then global methods are considered sufficient [20]. Both global and local approaches usually involve using a 2D target field with precisely known control point positions (or alternatively, accurately known inter-target distances) as a reference. The targets in the X-ray images can be selected semi-automatically [29] or fully-automatically [17]. The measured target positions are then compared to the reference positions and corrections are determined based on any differences. It is important to note that errors in the control points will cause a calibration error.

In the 3D space calibration step, Direct Linear Transform (DLT) or one of its variations is typically applied to solve for the geometric relationship (i.e. orientation and position) between the 3D target field (with accurately determined coordinates, e.g. from computed tomography, CT) and the fluoroscopic imaging system [22]. Although the DLT method has proven to be a simple and effective method for 3D space calibration, its accuracy is lower than more advanced techniques like bundle adjustment [30]. The accuracy of DLT can be improved by increasing the number of targets and ensuring the targets are evenly distributed [31], but DLT still undesirably calibrates each image independently. As described in [14], a calibration based on bundle adjustment constrains the IOP to be the same between all images originating from the same fluoroscope system, and can conveniently fix the ROP between two rigidly installed fluoroscope systems, which is not done in DLT. By treating all the images as having the same systematic errors and thus exploiting the redundancy in measurements and network geometry (i.e. reduced parameter correlation in the estimation), the accuracy and robustness of the calibration can be improved.

## III. Proposed Algorithm

Within the context of the pros and cons described in the previous section, the benefits of the proposed calibration method are highlighted below:

- Instead of using DLT, the fluoroscope is approximated as a pinhole camera, which allows the well-established photogrammetric bundle adjustment to be applied.

- Rather than requiring knowledge of the precise coordinates in the calibration target field (i.e. phantom), these

are treated as unknowns in the adjustment. This reduces the risk of fabrication or surveying errors propagating into the calibration solution, as well as potentially reducing the time and cost of manufacturing the target field.

- The object space coordinates, IOP, and EOP are estimated simultaneously to account for their correlations.

- Instead of relying on a high-order polynomial to model the systematic distortions identified in the literature review and the complex mapping of the actual fluoroscopy geometry to an idealized pinhole camera model, a k-nearest-neighbour regression is applied. In [14] and [26], the order of the global polynomial was manually selected based on an expert's judgement. In the proposed method, a K-fold cross-validation was used to select the optimal neighbourhood size (i.e. k) to reduce the subjectivity in the model selection [32]. According to [32], if the machine learning algorithm is stable for the given dataset, the algorithm should not be hypersensitive to the number of selected folds (i.e. K). From their experimentation with various open source datasets, they demonstrated that a K value between 10 and 20 typically yields a good bias-variance trade-off. The decision to use 10-fold cross-validation for model selection is generally also supported in literature by other authors; for example, [33] suggested using K=10 when the sample size is greater than 100 and [34] suggested using K=10 when the sample size is greater than 200. In this paper, 10-fold cross-validation was adopted with the smallest sample size being greater than 2000. Moreover, for the given dataset, experimentally increasing K to 20 did not have a statistically significant impact on the learned image distortion model. The adoption of kNN can describe local errors better than the global polynomial. Furthermore, the effect of unaccounted outliers will remain local in the calibration model; hence, only a small region of the calibrated image will be biased in the unfortunate event that an outlier has failed to be rejected.

- A robust Maximum Likelihood Estimation (MLE) based on the Student's t probability distribution is used instead of a least squares adjustment, whose L2-norm is known to be sensitive to outliers. This integrates the outlier detection/rejection step with the numerical optimization step and eliminates the need to perform post-adjustment blunder tests like Baarda's data snooping.

## IV. MATHEMATICAL MODEL

The proposed fluoroscope calibration algorithm is based on the concept of grey-box system identification [35]. While [36] have demonstrated that bundle adjustment can be solved as a black-box system using machine learning, it requires a lot of ground truth training data to learn the underlying geometry and it does not allow physical constraints to be enforced easily. On the other hand, the equations for projective geometry are well understood in the fields of computer vision and photogrammetry (i.e. they are considered a white-box model). The collinear relationship between the image measurements, pinhole camera position, and object space coordinates can be expressed compactly using a few parameters (1). These parameters can be determined accurately with just a few well distributed targets.

$$\vec{p}_{ijk} = \mu_{ijk} \times \vec{q}_{jk}(\vec{P}_i - \vec{T}_{jk})\vec{q}_{jk}^c \quad (1)$$

where, $\vec{p}_{ijk} = [x_{ijk} - x_{pk} - \Delta x_{ijk}, y_{ijk} - y_{pk} - \Delta y_{ijk}, -c_k]^T$ is the image coordinates of target $i$ in exposure $j$ captured by fluoroscopy system $k$. $x$ and $y$ are the image measurements, $x_p$ and $y_p$ are the principal point offsets, $c$ is the principal distance, and $\Delta x$ and $\Delta y$ are the image corrections.

$\vec{P}_i = [X_i, Y_i, Z_i]^T$ is the 3D object space coordinates of target $i$ on the phantom

$\vec{T}_{jk} = [X_{ojk}, Y_{ojk}, Z_{ojk}]^T$ is the 3D position of the fluoroscopy system $k$ relative to the phantom in exposure $j$

$\vec{q}_{jk}$ = unit quaternion representing 3D rotation of the fluoroscopy system $k$ relative to the phantom in exposure $j$. Superscript '$c$' represents the quaternion conjugate.

$\mu_{ijk}$ = unique scale factor of point $i$ in exposure $j$ of system $k$

To estimate these unknown parameters, the image distortion corrections, a.k.a. AP (i.e. $\Delta x_{ijk}$ and $\Delta y_{ijk}$) are first initialized to zero and treated as constants. The best set of IOP, EOP, and object space coordinates on the phantom can then be determined simultaneously using MLE. The statistically optimal point estimate can be rigorously estimated by minimizing the negative logarithm of the Student's t probability distribution (2). Unfortunately, the collinearity condition (1) is a non-linear functional model; therefore, it is first linearized using a first-order Taylor series expansion, which involves computing the Jacobian matrix analytically. Then the cost minimization step is done iteratively using the Levenberg–Marquardt algorithm until convergence [37]. The required gradient and Hessian information is computed analytically by taking the first and second derivatives of the negative logarithm of equation (2). Once converged, this point of minimum cost is equivalent to the point of maximum likelihood under the t-distribution. The rank deficiency in the normal matrix is resolved by defining the datum using inner constraints on the object space coordinates [38]. By using a minimally constrained network, the estimated image corrections are independent of the datum definition. The unknown IOP, EOP, and object space coordinates do not need to be estimated a priori with great accuracy, but their values should be initialized within the convergence region of the non-linear estimator to ensure convergence to the desired global minimum. After convergence, the image residuals and the variance-covariance matrix of the residuals are computed. Only the inliers are passed on to the next step to determine the AP.

$$F = \arg\max_{\vec{\theta}} \frac{\Gamma\left(\frac{v}{2} + \frac{D}{2}\right)}{\Gamma\left(\frac{v}{2}\right)} \frac{|C_l|^{-\frac{1}{2}}}{(v\pi)^{\frac{D}{2}}} \left[1 + \frac{\left(\vec{l} - f(\vec{\theta})\right)^T C_l^{-1} \left(\vec{l} - f(\vec{\theta})\right)}{v}\right]^{-\frac{v}{2} - \frac{D}{2}} \quad (2)$$

where, $\vec{l}$ is the vector of image measurements
$C_l$ is the variance-covariance matrix of the observations
$\vec{\theta}$ is the vector of unknown parameters

$f(\vec{\theta})$ represents the reprojected image coordinates
$v$ is the degrees-of-freedom
$D$ is the number of observations

The residuals from the bundle adjustment are composed of a deterministic part and a random part. The challenge is separating the two and describing the deterministic non-linear image distortions (i.e. $\Delta x_{ijk}$ and $\Delta y_{ijk}$) that arise from manufacturing limitations of each component and their alignment errors, in addition to errors from approximating the fluoroscope as a pinhole camera. Deriving a parametric functional model to describe this is non-trivial; therefore, a flexible non-parametric black-box model such as the kNN regression is used.

Nearby residuals in the image field are assumed to be spatially correlated because they are under the influence of a similar systematic error. In [13] the systematic error part is estimated by averaging 'k' neighbouring residuals. When many data points are available and are evenly distributed, this is a reasonable estimation. Although data acquisition is rather efficient and getting an abundance of data is becoming less of an issue with modern day technologies, it is still preferable to have an algorithm that can perform well using less data. Traditionally, kNN regression is known to be fast at training but slow at predicting, it typically produces a non-smooth surface, and in some cases, it performs worse than using the fixed-radius-nearest-neighbours scheme. In this paper, the first point is addressed by organizing the residuals using a KD-tree structure, which makes the predicting phase considerably faster. The second and third points are partly addressed by resampling the residuals uniformly over a rectilinear grid using linear interpolation (note: other grid-based resampling techniques can also be considered, such as cubic interpolation). To determine the optimal neighbourhood size, the hyper-parameter 'k' is determined using grid search with a 10-fold cross-validation to average the errors. The integer value 'k', which results in the minimum weighted L2-norm error metric is selected (3).

$$G = (\vec{r} - \vec{g}(x,y))^T C_r^{-1} (\vec{r} - \vec{g}(x,y)) \quad (3)$$

where, $\vec{r}$ is the vector of residuals from the bundle adjustment
$C_r$ is the variance matrix of the residuals (with only the diagonal components)
$x$ and $y$ are the cartesian coordinates in image space
$\vec{g}(x,y)$ is the vector of predicted residuals from the kNN regressor

Once the kNN regressor is trained, it is used to predict the magnitude and direction of the image distortion vector at every observed image point ($\delta x_{ijk}, \delta y_{ijk}$). Recall that $\Delta x_{ijk}$ and $\Delta y_{ijk}$ were initialized to zero at the beginning of the calibration, they can be updated as shown below:

$$\Delta x_{ijk} \mathrel{+}= \delta x_{ijk}$$
$$\Delta y_{ijk} \mathrel{+}= \delta y_{ijk}$$

With the updated $\Delta x_{ijk}$ and $\Delta y_{ijk}$ the bundle adjustment can be repeated to estimate a new set of IOP, EOP, and object space coordinates. This process of performing bundle adjustment with the AP set as constant, followed by doing kNN regression on the inlier residuals is then repeated until convergence. Convergence is achieved when the sum of the weighted cost function of the bundle adjustment (2) and the kNN regression (3) is minimized.

## V. EXPERIMENTATION

The same dataset as presented in [14] is used in this study; for more detailed information about the experiment, readers are advised to refer to the original paper. The following is a brief summary of the various components of the experiment.

### A. Imaging Systems

The two fluoroscopy systems used in this study are identical in terms of their architecture. Each consists of a rotating anode X-ray tube (Varian Medical Systems G-1086), high-voltage pulsed X-ray generators (EMD Technologies), image intensifier (Toshiba E5876SD-P2A), and high-speed digital video cameras (PCO AG DIMAX).

### B. Calibration Field

A 3D phantom with 503 radio-opaque spherical steel beads (3.5 mm diameter) distributed on four sides of an acrylic cube was used as the calibration target field. Although the accurate location of the beads is not required by the proposed calibration method, they were nonetheless surveyed using a coordinate measuring machine (FARO Technologies FaroArm) to serve as check points. The accuracy of individual measurements of the FaroArm is 0.025mm according to the manufacturer. Since it is impossible to directly measure the centroid of a steel bead, 15-25 measurements were acquired on the bead's surface and a geometric sphere fitting is performed to infer the centroid accurately.

### C. Data Acquisition

The calibration cube is rotated and translated to different locations within the field-of-view of the biplanar fluoroscopy system with the aid of a height-adjustable turntable. A total of 150 images per fluoroscopy system were acquired. Both fluoroscopy systems were installed on a stable base and assumed to be static during the entire experiment under a well-controlled laboratory environment. Furthermore, their image exposures are programmed to be time synchronized.

### D. Image Preprocessing

Rather than working with the high-dimensional raw radiographs, feature extraction is used to extract the center of the steel beads from the images for further processing. The target centroids were seeded and labelled semi-manually by the user and determined accurately to sub-pixel accuracy using 2D ellipse-fitting to the detected edge pixels. The centroid of the best-fit ellipse is then used as the input feature to the bundle adjustment. The average root mean-squared error (RMSE) of the geometric-form fitting was reported to be 0.06 +/- 0.01 pixels. This semi-manual centroid extraction took approximately 15 minutes for each image.

## VI. RESULTS AND ANALYSIS

Before processing the data, half the images are uniformly sampled and withheld for testing the efficacy of the proposed calibration. Therefore, each fluoroscopy system is only calibrated with a maximum of 75 images (i.e. 75 testing images and up to 75 training images per fluoroscope). The performance of the proposed calibration both in-sample (i.e. based on the training data) and out-of-sample (i.e. based on the testing data) are assessed by studying: 1) the image reprojection error from the bundle adjustment; 2) the mapping accuracy with the FaroArm measurements derived coordinates as the reference; and 3) the exterior orientation parameters with the expert photogrammetrist's solution using 150 images per fluoroscopy and calibrated together with relative orientation constraints as the reference. A 3D rigid body transformation was used to transform the 3D mapping coordinates and camera pose to the reference coordinate frame for the error analysis. To study the effect of the training sample size on the calibration, the number of training images used for each fluoroscopy system was varied between 15, 30, 45, 60, and 75. The effect of modelling the IOP explicitly in the bundle adjustment (rather than being implicitly learned from the kNN regressor) and smoothing the kNN regressor with linear resampling is studied. For each of these subsampled training images scenario, the calibration is performed in the following four schemes:

1. kNN: bundle adjustment estimates the object space coordinates and EOP, and the kNN regressor learns the image distortion and IOP.
2. kNN + IOP: bundle adjustment estimates the object space coordinates, EOP, and IOP, and the kNN regressor learns the image distortion.
3. kNN + smoothing: bundle adjustment estimates the object space coordinates and EOP, and the kNN regressor learns the image distortion and IOP after linear smoothing.
4. kNN + IOP + smoothing: bundle adjustment estimates the object space coordinates, EOP, and IOP, and the kNN regressor learns the image distortion after linear smoothing.

### A. Calibrating the fluoroscopy systems independently

The 2D and 3D in-sample errors of fluoroscope system #1 and #2 are presented in Figs. 1 and 2, respectively. The bundle adjustment results are analyzed by plotting the mapping errors, system position errors, and reprojection errors of the fluoroscopic system (on the y-axis) as a function of the number of training images (on the x-axis) under the different calibration schemes (in the legend).

In the absence of image distortion modelling (i.e. no calibration), both systems were able to achieve millimetre-level 3D mapping accuracy using 15 images when compared to the 3D reference centroids from the robotic arm (the average RMSE for fluoroscope #1 and #2 is 1.24 and 1.38 mm, respectively). This shows that the manufacturing quality is fairly consistent between the two systems (i.e. high manufacturing standards), with sub-millimetre discrepancies. In this experiment, fluoroscope #1 achieved a minimum 3D measurement error of 0.83 mm (33.2% improvement) when using 75 images instead of 15 images for 3D reconstruction. Likewise, fluoroscope #2 achieved a minimum error of 0.66 mm (52.2% improvement from using an additional 60 images). In general, increasing the amount of training data (even without doing calibration, i.e. only solving for the EOP and object space coordinates without modelling the image distortions) improves the 3D mapping accuracy because of the increased redundancy.

However, the same cannot be said about the 2D reprojection error and 3D camera position error. In fact, increasing the number of images from 15 to 75 increased the average reprojection errors from 1.09 to 1.30 pixels (for fluoroscope #1), and from 1.20 to 1.30 pixels (for fluoroscope #2). The 3D camera position accuracy initially improved with increasing number of images (e.g. from 15 to 45), but ultimately the accuracy deteriorates with additional images (i.e. 75 images). These two findings can be attributed to the unmodelled systematic errors. While the free-network bundle adjustment tends to give an accurate 3D reconstruction, due to projective compensation, the errors are propagated into the EOP and therefore image reprojection.

Modelling the image distortions improves the 3D reconstruction accuracy by greater than 80% in all calibration cases. The 2D reprojection error reduced from pixel-level to less than ⅓ of a pixel in all calibrations for both fluoroscopes. The most effective calibration scheme appears to be using either kNN + smoothing or kNN + IOP + smoothing. In the best-case scenario with 75 training images, the average 3D mapping error reduced from 0.83 mm to 0.04 mm for fluoroscope #1 (95.5% improvement), and from 0.66 mm to 0.04 mm for fluoroscope #2 (94.4% improvement). The corresponding 2D reprojection error reduced from 1.17 to 0.17 pixels for fluoroscope #1 (85.8% improvement), and 1.30 to 0.18 pixels for fluoroscope #2 (85.9% improvement). After calibration, the geometric measurement performance of the two fluoroscopic imaging systems became more comparable; the discrepancy between the two systems was reduced by an order of magnitude in terms of mapping accuracy (~0.1 to ~0.01 mm) and reprojection error (~0.1 to ~0.01 pixel).

Estimating the IOP explicitly in the bundle adjustment versus grouping them with the AP in the kNN is more controversial as the benefits are not appreciated in all situations. Compared to the fluoroscope positions estimated by the reference bundle adjustment solution, estimating the IOP explicitly in the proposed adjustment makes the fluoroscope position estimates more consistent in all cases for fluoroscope #1. For fluoroscope #2, some variations to this error pattern can be observed. This also suggests that errors in the EOP may not be a good surrogate measure of the quality of the calibration. Part of the EOP error can be attributed to the accuracy of the reference camera pose solution, which may not be an order of magnitude better than the proposed solution as would normally be desired in a check point analysis. Furthermore, due to the correlations that exist between the EOP, IOP, and AP, the EOP may appear different from the reference solution, but when combined with a different IOP and AP, it results in an accurate reconstruction in 3D and projection in 2D. This phenomenon, often referred to as projective compensation, is a well-known issue in photogrammetry. It is included here to demonstrate that accurate fluoroscope positioning is challenging for such narrow field-of-view imaging systems; but nevertheless, the proposed calibration appears to be able to improve the overall geometry of the solution. For the remainder of this paper, the EOP will not be analyzed any further due to projective compensation.

Despite the difference in scale, the reprojection error bar graph, Fig. 1 (e), shows a similar trend as the mapping errors plot, Fig. 1 (a). More data is usually desirable from an accuracy perspective, up to a certain limit. Applying a smoothing function shows a noticeable benefit in the reduction of fluoroscope positioning errors, mapping errors, and reprojection errors. Additionally, the smoothed kNN regressor makes the calibration less sensitive to sample size. For both fluoroscopes, when looking at the reprojection errors and mapping errors, it is difficult to appreciate the benefit of additional images beyond using 45 images (i.e. a redundancy of 13.1 thousand); the reprojection error remains unchanged and the mapping errors in fluoroscope #1 and #2 reduce by an additional 0.007 mm and 0.006 mm with an extra 30 images, respectively, which is insignificant considering the measurement tolerance of the FaroArm.

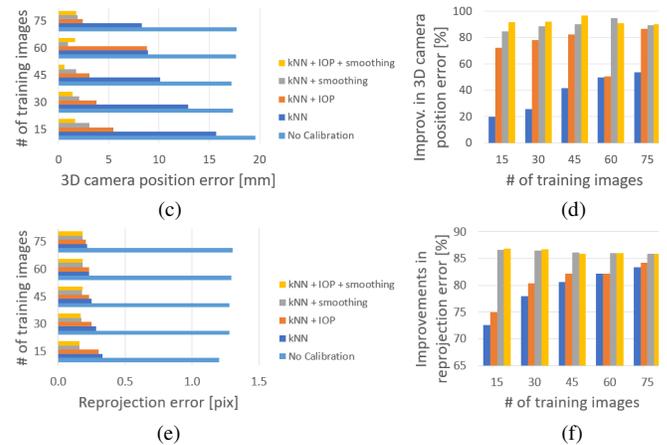

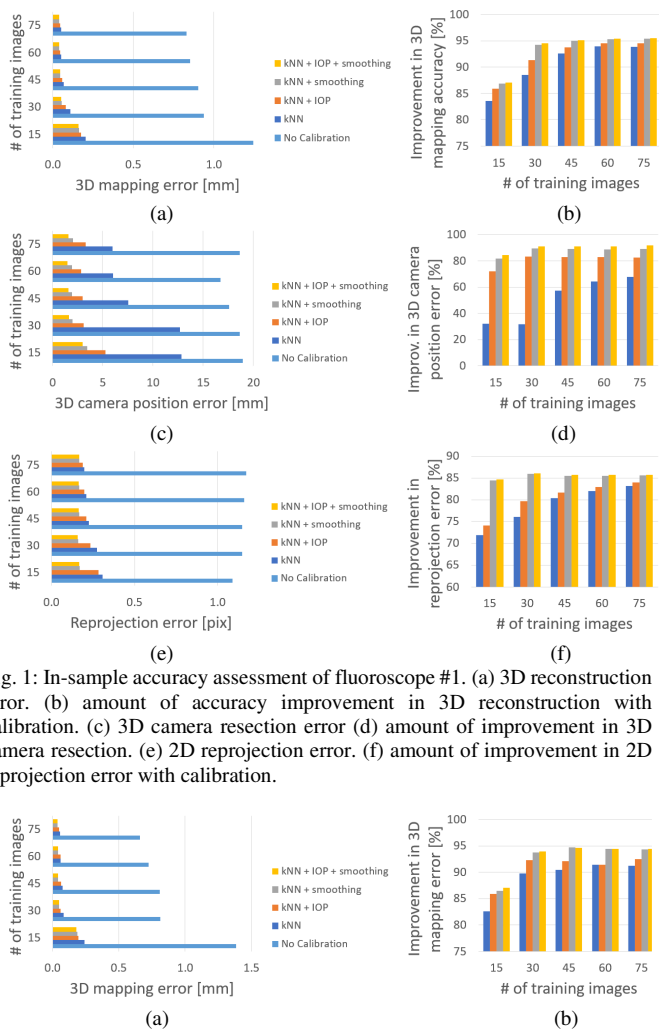

Fig. 1: In-sample accuracy assessment of fluoroscope #1. (a) 3D reconstruction error. (b) amount of accuracy improvement in 3D reconstruction with calibration. (c) 3D camera resection error (d) amount of improvement in 3D camera resection. (e) 2D reprojection error. (f) amount of improvement in 2D reprojection error with calibration.

Fig. 2: In-sample accuracy assessment of fluoroscope #2. (a) 3D reconstruction error. (b) amount of accuracy improvement in 3D reconstruction with calibration. (c) 3D camera resection error (d) amount of improvement in 3D camera resection. (e) 2D reprojection error. (f) amount of improvement in 2D reprojection error with calibration.

While it is important to assess the quality of the data and results used in the calibration, it is useful to decouple the calibration quality assessment from the data used for training (due to the undesirable effect of projective compensation). Applying the calibration learned from the training dataset on the testing dataset gave the results in Fig. 3 and Fig. 4 for fluoroscopes #1 and #2, respectively. Unlike Fig. 1 and Fig. 2, where the errors were calculated using varying numbers of images, 75 training images were consistently used for quality control. This ensures that any decreases in mapping error and reprojection error are a result of the image correction.

It can be perceived in Fig. 3 and Fig. 4 that kNN + IOP usually outperforms kNN alone, and smoothing the kNN results can yield lower errors than kNN + IOP. Once smoothing is performed, whether the IOP are explicitly modelled in the bundle adjustment or are learned implicitly by the kNN regressor does not yield a noticeable difference. In the absence of AP modelling, the 3D measurement error is 0.83 and 0.61 mm with a corresponding 2D image reprojection error of 1.17 and 1.31 pixels, respectively for fluoroscopes #1 and #2. After error modelling using a smoothed kNN, for fluoroscope #1, the 3D measurement error ranges from 0.04—0.05 mm (i.e. 93.5—95.7% improvement) with a 2D reprojection error ranging from 0.20—0.29 pixels (i.e. 75.1—83.2% improvement), depending on the amount of training data used. Similarly, for fluoroscope #2, the 3D measurement error is approximately 0.04 mm (i.e. 93.3—94.2% improvement) with a 2D reprojection error ranging from 0.21—0.29 pixels (i.e. 77.8—83.8% improvement). This appears to be consistent with the pattern from the in-sample errors above.

When combined with a smoothing function, the kNN regressor can be trained with approximately 45 images, as that is when the error curve begins to plateau (see Fig. 3 and Fig. 4). Without the smoothing function, more images can be considered in order to improve the geometric accuracy. However, even doubling the training data for the conventional kNN regressor hardly achieves the same level of accuracy as the smoothed kNN regressor. Modelling the errors with a smoothed kNN regressor also results in more rapid convergence as demonstrated in Fig. 5. Regardless of the calibration scheme, the adjustment typically converges in 30 iterations. The C++ program was tested on a laptop running a single core 2.80 GHz CPU with 8 GB of RAM, with each iteration taking between 8.1 sec (for 15 images) to 42.1 sec (for 75 images).

Fig. 6 and Fig. 7 compare the adjustment residuals and the estimated distortion model using kNN regression and smoothed kNN regression using 15 and 75 training images. It is demonstrated that the proposed calibration can reduce the magnitude of the overall residuals and result in a more homogenous appearance. Differences between kNN and smoothed kNN regression can be seen in all cases, albeit having more data undoubtedly made the conventional kNN regression solution smoother looking, the final learned systematic distortion model is different. Furthermore, the smoothed kNN also appears to be less dependent on sample size.

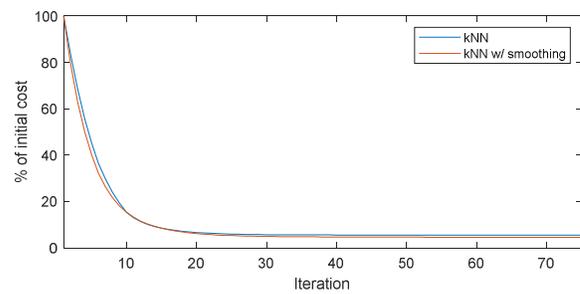

Fig. 5: comparison of the convergence rate between using kNN regressor and smoothed kNN regressor for modelling the AP

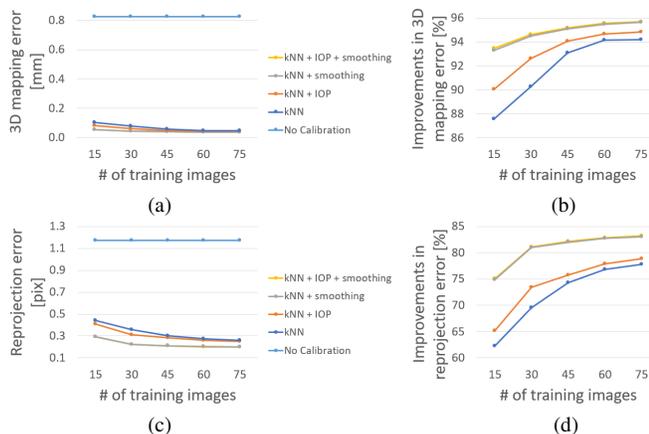

Fig. 3: Out-of-sample accuracy assessment of fluoroscope #1. (a) 3D reconstruction error. (b) amount of accuracy improvement in 3D reconstruction with calibration. (c) 2D reprojection error. (d) amount of improvement in 2D reprojection error with calibration.

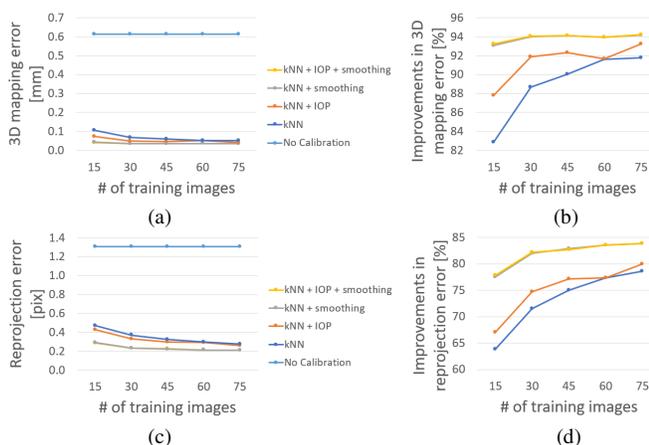

Fig. 4: Out-of-sample accuracy assessment of fluoroscope #2. (a) 3D reconstruction error. (b) amount of accuracy improvement in 3D reconstruction with calibration. (c) 2D reprojection error. (d) amount of improvement in 2D reprojection error with calibration.

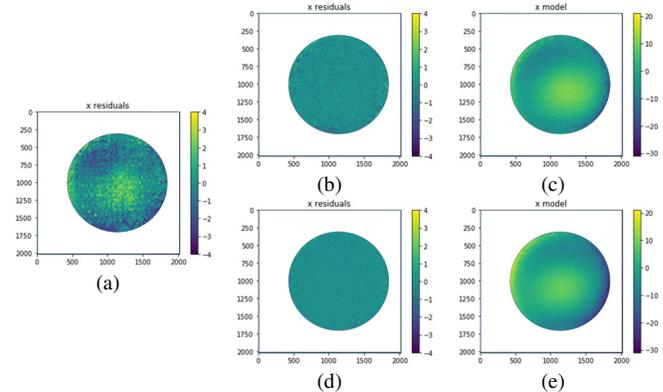

Fig. 6: Systematic and random components of x-residuals from training of fluoroscope #1 using 15 images. (a) residuals measured in pixels after bundle adjustment without calibration. (b) and (d) show the residuals after calibration using kNN and smoothed kNN regression, respectively. (c) and (e) are the learned distortion model using kNN and smoothed kNN regression, respectively.

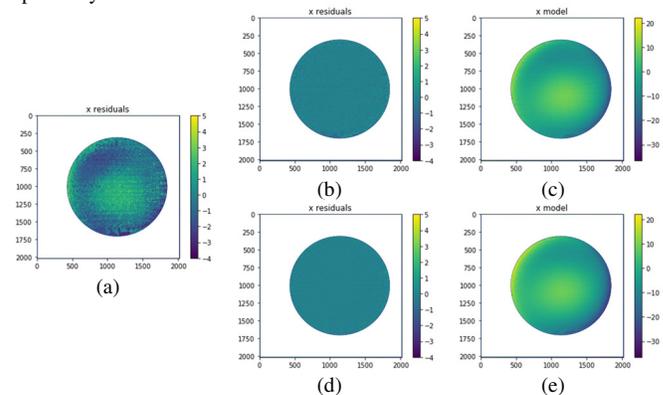

Fig. 7: Systematic and random components of x-residuals from training of fluoroscope #1 using 75 images. (a) residuals measured in pixels after bundle adjustment without calibration. (b) and (d) show the residuals after calibration using kNN and smoothed kNN regression, respectively. (c) and (e) are the learned distortion model using kNN and smoothed kNN regression, respectively.

### B. Calibrating the fluoroscopy systems jointly with relative orientation constraint

When jointly calibrating two rigidly mounted fluoroscopes, a geometric constraint can be easily applied to the bundle adjustment to enforce a fixed rotation and translation between the two systems. This can potentially improve the calibration results, although the effect is small in this case. Using all training images, the in-sample 3D mapping accuracy before and after error modelling is 0.62 and 0.03 mm, respectively (i.e. 94.6% improvement). Fig. 8 graphically shows the in-sample errors from the dual-fluoroscopy calibration, which are comparable to the errors from when the systems were calibrated separately, except with a dual-fluoroscope system, more images are captured at each epoch (two images are captured instead of one). Therefore, the errors plateau even earlier; with about 30 image pairs, the gain in calibration accuracy has already levelled off (i.e. redundancy of 17.7 thousand). The reprojection error reported here using the proposed calibration is approximately 0.17 pix, which is less than the 0.45 pix error reported in [14]. The improvement is likely attributable to the local distortion modelling achieved using the kNN regression.

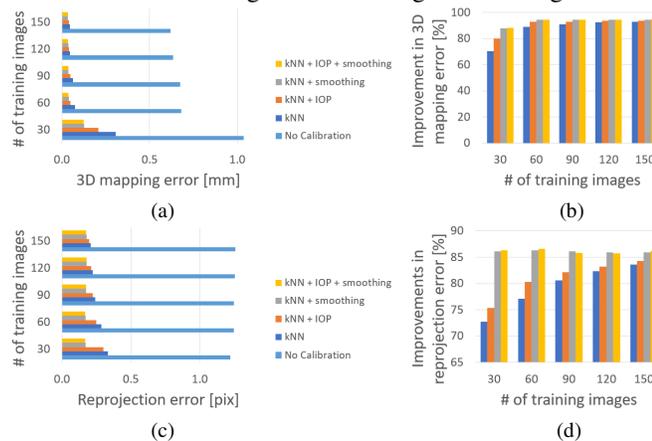

Fig. 8: In-sample accuracy assessment of biplanar fluoroscopy. (a) 3D reconstruction error. (b) amount of accuracy improvement in 3D reconstruction with calibration. (c) 2D reprojection error. (e) amount of improvement in 2D reprojection error with calibration.

During an endovascular procedure, it may not be possible to acquire multi-view radiographs and perform 3D reconstruction using bundle adjustment. One of the main benefits of using a dual-fluoroscope instead of a single-fluoroscope is the ability to perform two-ray spatial intersection to infer the 3D information with a metric scale in closed-form once the corresponding points have been identified. This is more suitable for analyzing dynamic scenes and recognizing temporal patterns. In light of this, to analyze the out-of-sample errors for a dual-fluoroscopy system, the 3D positions of the steel beads were first determined using spatial intersection from pairs of radiographs, and then they were transformed to the coordinate system of the mapping frame using a 3D rigid body transformation. The average errors relative to the reference coordinates are then computed (Fig. 9 and Table I). Without modelling the AP, or no calibration (i.e. only solving for the ROP), 3D mapping errors are between 0.59—0.68 mm, and after AP modelling using a smoothed kNN, the mapping errors are between 0.31—0.32 mm (i.e. 46.2—53.8% improvement). This 3D reconstruction accuracy is comparable to the results from [14], where they reported an error between 0.3–0.4 mm; however, it is worth noting that a different and larger set of testing images was used in this paper. In addition, when a smoothed kNN regressor is used for error modelling, as little as 15 pairs of images (i.e. redundancy of 8.1 thousand) is sufficient to achieve accuracies in the 0.3 mm range; extra training images had little impact on the system's overall geometric accuracy. This is significantly less than 150 image pairs used in [14], where a global parametric model was used for modelling the AP. In addition, it shows that by introducing the ROP, the geometric network is strengthened and less data is needed for calibration.

TABLE I: AVERAGE TWO-RAY SPATIAL INTERSECTION ACCURACY OF BIPLANAR FLUOROSCOPY BEFORE AND AFTER ERROR MODELLING. NOTE: A RELATIVE ORIENTATION AND POSITION CONSTRAINTS HAVE BEEN APPLIED IN ALL CASES.

| # of image pairs | Calibration Mode | $X_{RMSE}$ [mm] | $Y_{RMSE}$ [mm] | $Z_{RMSE}$ [mm] | Average RMSE [mm] | % Improvement |
|---|---|---|---|---|---|---|
| 15 | No calibration | 0.76 | 0.47 | 0.76 | 0.68 | N/A |
|  | kNN | 0.26 | 0.35 | 0.35 | 0.32 | 52.8 |
|  | kNN + IOP | 0.27 | 0.35 | 0.36 | 0.33 | 51.6 |
|  | kNN + smoothing | 0.24 | 0.35 | 0.34 | 0.31 | 53.8 |
|  | kNN + IOP + smoothing | 0.24 | 0.35 | 0.35 | 0.31 | 53.8 |
| 30 | No calibration | 0.63 | 0.46 | 0.65 | 0.59 | N/A |
|  | kNN | 0.25 | 0.34 | 0.35 | 0.31 | 46.3 |
|  | kNN + IOP | 0.25 | 0.34 | 0.35 | 0.32 | 45.9 |
|  | kNN + smoothing | 0.24 | 0.34 | 0.35 | 0.31 | 46.3 |
|  | kNN + IOP + smoothing | 0.24 | 0.34 | 0.35 | 0.32 | 46.2 |
| 45 | No calibration | 0.63 | 0.45 | 0.66 | 0.59 | N/A |
|  | kNN | 0.24 | 0.34 | 0.35 | 0.31 | 46.5 |
|  | kNN + IOP | 0.25 | 0.34 | 0.36 | 0.32 | 46.0 |
|  | kNN + smoothing | 0.24 | 0.34 | 0.35 | 0.31 | 46.4 |
|  | kNN + IOP + smoothing | 0.24 | 0.34 | 0.35 | 0.32 | 46.3 |
| 60 | No calibration | 0.64 | 0.45 | 0.66 | 0.59 | N/A |
|  | kNN | 0.24 | 0.34 | 0.35 | 0.31 | 47.0 |
|  | kNN + IOP | 0.25 | 0.34 | 0.36 | 0.32 | 46.7 |
|  | kNN + smoothing | 0.24 | 0.34 | 0.35 | 0.31 | 46.9 |
|  | kNN + IOP + smoothing | 0.24 | 0.34 | 0.35 | 0.32 | 46.8 |
| 75 | No calibration | 0.65 | 0.45 | 0.67 | 0.60 | N/A |
|  | kNN | 0.25 | 0.34 | 0.35 | 0.32 | 47.2 |
|  | kNN + IOP | 0.25 | 0.34 | 0.35 | 0.32 | 47.1 |
|  | kNN + smoothing | 0.24 | 0.34 | 0.35 | 0.32 | 47.0 |
|  | kNN + IOP + smoothing | 0.24 | 0.34 | 0.35 | 0.32 | 47.2 |

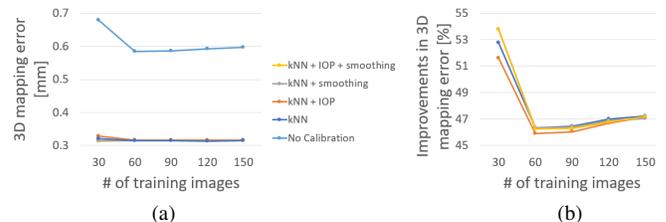

Fig. 9: Out-of-sample accuracy assessment of biplanar fluoroscopy, i.e. Fluoroscope #1 and #2 together with relative pose constraint. (a) 3D reconstruction error. (b) amount of accuracy improvement in 3D reconstruction with calibration.

## VII. CONCLUSION

Fluoroscopy is an important biomedical imaging modality in modern day medicine. It allows interventional radiologists and vascular surgeons to perform minimally-invasive endovascular surgery accurately that would otherwise be impossible. As the interventionist cannot visualize the catheter directly, they have to rely on the fluoroscopic data. Undoubtedly, any errors in the images will lead to errors in the catheter insertion/navigation. This will be particularly true in the future when endovascular surgery will become more robot-assisted. This paper presented a data-driven robust automatic self-calibration approach based on photogrammetric bundle adjustment and smoothed kNN regression to model the systematic image distortions.

This study demonstrated that using modern day fluoroscopy, 3D multi-view object space measurement errors ranging from 0.61—0.83 mm and 2D image measurement errors ranging from 1.17—1.31 pixels are attainable before calibration. After the proposed photogrammetric bundle adjustment with self-calibration procedure using a self-tuned smoothed kNN regressor, these errors are reduced to 0.04—0.05 mm (i.e. 82.9%—95.7% improvement) and 0.20—0.29 pixels (i.e. 93.3%—95.7% improvement). The most significant improvement in accuracy occurs with the first 45 training images (or a redundancy of approximately 13.1 thousand); beyond that, adding more images showed only a mild improvement.

In the case of a dual-fluoroscopy setup, the 3D measurement accuracy plateaued with approximately 15 image pairs (or a redundancy of 8.1 thousand); at this point the 3D reconstruction error was reduced from 0.68 to 0.31 mm (i.e. 53.8% accuracy improvement) and adding additional training data did not improve the calibration significantly. Less images are required compared to the multi-view case partly because a relative orientation constraint can be enforced. In this setup, the 3D accuracy is lower because the depth is only recovered from the spatial intersection of two conjugate light rays; however, these conditions are more indicative of its application in clinical practice for real-time 3D motion tracking. When compared to the previous method where an expert photogrammetrist had to manually adapt the model to the data [14], the proposed self-supervised learning calibration method can achieve similar geometric accuracies without manual tuning and using less data. In general, a smoothed kNN outperformed a standalone kNN regressor for image distortion modelling and is strongly recommended.

## VIII. ACKNOWLEDGEMENT

Special thanks to Kathleen Ang for proof-reading and editing this paper.